\documentclass[12pt,preprint]{aastex}
\def\vect#1{\stackrel{\rightarrow}{#1}}
\begin{document}

\title{Competition Between Gravitational and Scalar Field Radiation}

\author{Demosthenes Kazanas\altaffilmark{1}
 \& Vigdor L. Teplitz\altaffilmark{2}}
\affil{Laboratory for High Energy Astrophysics, NASA Goddard Space Flight
Center,
Code 661, Greenbelt, MD 20771, USA.}
\altaffiltext{1}{email: Demos.Kazanas-1@nasa.gov}
\altaffiltext{2}{email: Vigdor.L.Teplitz@nasa.gov; Permanent address: Physics Department, Southern Methodist University, Dallas, TX 75725}

%\authoraddr{}
%\email{}

\begin{abstract}
Recent astrophysical observations have provided strong evidence 
that the present expansion of the universe is accelerating, powered
by the energy density associated with a cosmological term. Assuming 
the latter to be not simply a constant term but a ``quintessence" 
field, we study the radiation of quanta of such a ``quintessence" field 
(``quintons") by binary systems of different types and compare intensities 
to those of standard tensor gravitational wave emission. We consider 
both the case in which the quintessence field varies only over cosmological 
distances and the case in which it is modified spatially by (strong) 
gravitational fields, a condition that results in bounds on the gradient 
of the scalar field. We show that, in both the first case and, because 
of a bound we derive from the Hulse-Taylor pulsar, in the second, there is 
not sufficient quinton radiation to affect expected LISA and LIGO gravity 
wave signals  from binary systems.  We show that, in the second case, 
the Large Hadron Collider is capable of setting a bound similar to 
that from the binary pulsar.

\end{abstract}


\section{Introduction}

The past few years have witnessed a revolution in cosmology, a field 
that has been growing from data poor to data rich by leaps and bounds
since the early 80's. This revolution followed the discovery, 
by the study of distant supernovae (Garnavich et al. 1998, 
Perlmutter et al. 1998), that the expansion of the universe is currently accelerating. A similar conclusion was reached in the ``concordance 
analysis" of the cosmic microwave background (CMB) data by the WMAP 
collaboration (Bennett et al. 2003; Page et al. 2003) which concluded 
that the universe  has a value of $\Omega$, the ratio of its density 
to the critical density, consistent with $\Omega =1$.  
In addition, the analysis of the WMAP data allowed an independent 
estimate of the contribution to this value by matter  alone (including 
that of the ``dark matter" which makes up most of the gravitationally 
bound matter); this was found not to exceed 30\%, suggesting that 
a cosmological constant term could be responsible for the remainder, 
leading to an estimate $\Omega_{\Lambda} \simeq 0.7$.

The presence of a cosmological term of this magnitude has been
the cause for great concern in the field: The ``natural" value of 
such a term in the context of theories of the fundamental interactions 
would be huge, of order of the Planck energy density, the  only 
``typical" value in a theory that contains a mass scale such as the 
Planck mass $M_P$ (Carroll 2000; Peebles \& Ratra 2003). 
The hope, then, was that a (hitherto 
unknown) conservation law would set its value to precisely zero, a value 
generally consistent with the pre-1998 data. The presence of a small, 
non-zero value for the cosmological term is then seen as another 
``fine tunning" problem in cosmology. Nevertheless, 
independently of the issue of the magnitude of the cosmological term, 
it is generally expected that this term does not represent a universal
constant associated with the Lagrangian of gravitational interaction. 
Rather, it is thought that it likely varies like a field, whose value 
is constant on cosmological scales, while its magnitude is varying slowly 
in time. This field, being quite distinct from the other known fields 
and forms of matter has been given the name quintessence (Caldwell, Dave 
\& Steinhardt 1998), borrowing the nomenclature from Aristotle's ``fifth 
substance" that was supposedly involved in the make up of the universe.

Once, however, one decides that the term that drives the acceleration 
of the universe is not an {\sl a priori} constant but a (scalar) field, 
one is immediately led to the notion that such a field obeys its own
(scalar wave) equation and that it should be neither constant in time
(a feature generally used in the literature) nor uniform on
non-cosmological scales. As such it carries its own ``kinetic"
energy and potential energy which couple to the gravitational
field thereby affecting both the metric around a bound object and also
being affected by it, through the contributions  of the metric to 
the covariant derivative.

Interestingly enough, an exact solution of the combined set of Einstein
and zero-mass scalar field equations in the static spherically symmetric 
case has been derived independently several times in the literature,
(e.g. Buchdahl 1959) including by one of us (Mannheim \& Kazanas 
1991 ). This last reference in particular, conjectured that the 
scalar field considered is none other than the Higgs field of high 
energy physics, which presumably is responsible for giving masses 
to particles through a spontaneous symmetry breaking process. 
This is rather relevant in that our present study involves also 
the study of its effects in high energy collisions  (of 
course, this field is expected to make the usual (huge) contribution
to the cosmological constant, which must be somehow cancelled as 
discussed above; the effects discussed by Mannheim \& Kazanas (1991) 
pertain only to the effects of the kinetic energy of the 
scalar field).

Motivated by the above considerations, we examine, in the present note,
the case in which both the scalar and gravitational fields are space 
and time dependent, i.e. the case of quinton emission by their 
combined action. Such a study becomes imperative with the development 
of detectors of gravitational radiation that either currently
are (TAMA) or about to become operational (LIGO), or are planned to
be built in the not too distant future (LISA).  Identifying 
gravitational waves in these facilities will depend on the comparison
of the data to theoretically-derived, computer-generated, wave forms.
These forms, and hence the interpretation of the data, could be 
impacted by competing processes not accounted for in template
generation.

Due to the nature of our study, we have proceeded classically in the 
linearized regime with our main goal seeking conditions and physical 
situations for which radiation of scalars might become comparable to 
that of gravitational waves. To this end we explore both astrophysical 
systems, as well as systems in the laboratory (e.g. high energy 
collisions).  We will also explore both the case in which the variation 
of the scalar field is limited to cosmological times and distances
and the case in which the scalar field configuration is influenced by 
the presence of nearby (compact) objects.

Since we will be comparing scalar radiation with graviton radiation, 
we recall here (Shapiro \& Teukolsky 1983) classical order of magnitude 
formulae for the latter.  For a binary system, gravitational radiation
luminosity is given in terms of the reduced and total masses, $\mu$ and
$M$ respectively, by the expression
\begin{equation}
    L_{GW} = (32/5)G^4M^3\mu^2/(c^5a^5)
\end{equation}
with $M=M_1+M_2$, $\mu^{-1} = M_1^{-1} + M_2^{-1},$ and $a = a_1 + a_2$
where $M_i$ and $a_i$ are the mass and distance from the center of mass (and
origin) of body $i~ (i=1,2)$  The rate at which the period, $\nu^{-1} = P =
2\pi/\omega$ $(\omega^2a^3 = GM)$, changes is given by
\begin{equation}
    P^{-1}dP/dt = -(96G^3/5c^5)M^2\mu/a^4
\end{equation}

In \S 2 we set out the relevant equations and their linearized form
while in \S 3 we examine the intensity of scalar field radiation 
resulting from the changing gravitational field (or metric) of a 
binary system and compare it to that of the gravitational radiation.  
We work, first, within the approximation that the (cosmological) scalar
field is not affected by the gravitational field of the binary system. 
In \S 4 we examine the competition between gravitational and
scalar radiation taking into account the change in the scalar field due to the
gravitational field of the binary using the solution of Mannheim \& 
Kazanas (1991) and finally, in \S 5 we present and discuss our conclusions.

\section{Basic Equations}

Let $\Psi$ denote a scalar field obeying the scalar field equation
\begin{equation}
\Box \Psi = (\partial_t^2 - \nabla^2)\Psi -
\Gamma^{\mu} \partial_{\mu}\Psi = -m_{\rm eff}^2 \Psi =
-\frac{\partial V(\Psi)}{\partial \Psi}
\label{box}
\end{equation}
where the box operator denotes the covariant d' Alembertian, $\Gamma^{\mu}$
are the Christoffel symbols associated with the (in general) curved
space into which $\Psi$ operates and $m$ is the (effective) mass of the
(quinton) field $\Psi$.  Our calculations are based on the approximation of first
solving Equation (\ref{box}) for the ``semi-static'' case and then computing the
radiation field by approximating the right hand side by zero and using the
$\Gamma$-term, with the semi-static quinton field, as the quinton radiation
source.

Assuming space--time to be sufficiently close to flat, one can use
a perturbative approach to calculate the components of the metric
tenson $g^{\mu \nu}$ and then the Christoffel symbols. Thus, in the slow velocity regime, one can expand the
metric tensor to powers of $v/c$, where $v$ is the magnitude of the
velocity of the matter components involved. In this approximation,
to second order in $v/c$ in $g^{00}$ and to first order in $v/c$
in $g^{ii}$, the space part of the metric tensor, the departure from
flat is equal to twice the gravitational potential $\phi$. In
this approximation the diagonal metric tensor components and the
square root of its determinant are (Weinberg 1972), remebering that
$-g = {\rm det}(g_{\mu \nu})$,
\begin{equation}
g^{\mu \nu} = [g^{00}, g^{ii}]= [(-1+2\phi), (1+2\phi)], ~~{\rm and} ~~
(-g)^{1/2} = \left[(1-2 \phi)^2 \right]^{1/2} = 1-2 \phi +
{\cal O}(\phi^2)
\end{equation}
leading to
\begin{equation}
(-g)^{1/2}g^{\mu \nu} = \left[-1+4\phi+ {\cal O}(\phi^2), 1+{\cal O}
({\phi^2})\right]
\end{equation}
which then yields for the Christoffel symbols
\begin{equation}
\Gamma^{\mu} = (-g)^{-1/2}\partial_{\nu}\left[(-g)^{1/2}g^{\mu \nu}\right] =
\left[\partial_0 \phi, {\cal O}(\phi^2)\right]
\end{equation}
With the above expression for $\Gamma^{\mu}$, the equation obeyed by
the scalar field in curved space--time becomes
\begin{equation}
(\partial_t^2 - \nabla^2)\Psi - 4 \, \partial_0 \phi
\partial_0 \Psi = - m^2 \Psi \simeq 0
\end{equation}

This equation is a wave equation with an inhomogeneous term. Its
solution can be found in standard texts (e.g. Jackson 1962), i.e.
\begin{equation}
\Psi_R(x, t) = 4\int \frac{d^3 x^{\prime}}{\vert
\stackrel{\rightarrow}{x} - \vect{x'}\vert}
\partial_0 \Psi_0(\stackrel{\rightarrow}{x'},t') \partial_0 \phi(\vect{x'}, t')
\end{equation}
where $\Psi_0$ is the unperturbed (cosmologically varying only) scalar
field at point $x'$ and time $t'$ and we are suppressing factors of $c$
except when evaluating expressions numerically.

Considering the combined gravitational -- scalar field variations appropriate
for a binary system in circular orbit centered on the center of mass of
the two bodies, the above equation reduces to
\begin{equation}
\Psi_R(x, t) = 4G (\partial_0 \Psi_0)
\int \frac{d^3 x'}{\vert \vec{x'} - \vec{x} \vert}  \left[M_1\frac{\vec{x'} - \vec{a_1}(t_1)}{\vert \vec{x'} - \vec{a_1}(t_1)\vert^3 } \cdot \dot {\vec a}_1(t_1)
 + M_2\frac{\vec{x'} - \vec{a_2}(t_2)}{\vert \vec{x'} -
\vec{a_2}(t_2)\vert^3 } \cdot 	\dot {\vec a}_2(t_2) \right]
\label{psi}
\end{equation}

where
\begin{equation}
t' = t - \vert \vec{x} - \vec{x'} \vert  \simeq t - x + \hat{x}\cdot
\vec{x'}
\end{equation}
and
\begin{equation}
t_i = t' - \vert \vec{x'} - \vec{a_i}(t_i) \vert \simeq t' - x' + \hat{x'}\cdot
\vec{a_i}(t_i)
\end{equation}
where $\vec{a_i}(t'_i)$ is the ``doubly retarded" vector connecting the 
center of mass of the system with body $i$. In other words $t_i$ is the 
time at which a light signal from $\vec{a}_i(t_i)$ would have to leave 
in order to get to $\vec{x}$ at time
$t$  via  $\vec{x'}$ at time $t' = t - \vert \vec{x} - \vec{x'} \vert$.  
We will also use $\partial_0\Psi = m_{\rm eff}\Psi_0$, with $m_{\rm 
eff}^2\Psi_0^2 \simeq U_{DE}$, where $U_{DE} = 10^3eV/cm^3$ is the 
observed dark energy density   The quinton (scalar field) at $\vec{x}$ 
is the result of coherent addition of the $\Psi$ generated by the time 
dependent field at $\vec{x'}$ from the masses at $\vec{a}_i(t_i)$.

In the long wavelength approximation ($\omega^{-1}>a$), most of the 
contribution to the integral comes from distances less than $x' \sim 
\pi/\omega$. We  take $\vert \vec{x} - \vec{x'} \vert \simeq x$ for the
outer denominator.  We use
\begin{equation}
\vect{a_i}(t_i) = \vect{a_i}(t'-\vert\hat{x'}-\vect{a_i}(t_i)\vert) \simeq \,
\vect{a_i} e^{i\omega (t'-\vert\hat{x'}-\vect{a_i}(t_i)\vert)} \simeq \,
\vect{a_i} e^{i\omega (t - x)} (1+i\omega \hat{x'}\cdot\vect{a_i})
\end{equation}
In the same (long wavelength) approximation, we can ignore differences 
between the time dependence of $a_1$  and $a_2$ beyond that included in 
Equation (12).  We now perform the time derivatives of the $\vect{a}$' s 
in Equation (9) making use of Equation (12).  The result is
\begin{equation}
\Psi_R(\vect{x},t) = 4GU_{DE}^{1/2} e^{2i\omega(t-x)} x^{-1} \int dx' 
d\Omega' \{ \omega^2 \, \hat{x'} \,[(\hat{x'}\cdot \vect{a_1})^2M_1 + 
(\hat{x'}\cdot \vect{a_2})^2M_2] 
%- i \omega [\vect{a_1}^2 M_1 + \vect{a_2}^2 M_2] 
\}
\end{equation}

Other terms, at least in lowest order approximation, vanish and/or have the wrong time dependence.  We make the approximations: (1) that the lower limits in the two $x'$ integrals, $a_1$ and $a_2$ are both approximately $a/2$, with $a = a_1+a_2$ and (2) that the upper limits are of order $\pi/\omega$ past which the oscillating exponentials dampen any contribution. We use the fact that $M_1a_1^2+M_2a_2^2 = \mu a^2$ to obtain
\begin{equation}
\Psi_R(x,t) = (16\pi^2/3x) e^{2i\omega(t-x)}\, G \,U_{DE}^{1/2} a^2\omega\mu
\end{equation}
With this result, the fact that the intensity $L_{\Psi}$ of quinton 
radiation is $4\pi x^2T^{0i}$, and the expression for the $0i$-component 
of the scalar field  stress-energy tensor $T^{0i} = \partial_t \, \Psi_R \,
\partial_r \Psi_R$, we obtain for the intensity
\begin{equation}
L_{\Psi} \simeq 10^4 U_{DE} \,  G^2 \mu^2 \, a^4 \omega^4
\label{Lpsi}
\end{equation}
where we have integrated over $\theta$ and averaged over $\theta_{\Psi}$.
This expression is applied to several specific cases in the next section.
And, in the section after that, the analogue to this result, for the case in
which the scalar field solution is modified by the gravitational field
of the compact system, is derived and applied.

\section{Slowly Varying Quintessence Field}

We consider the case in which the scalar (quintessence) field (in
section II) is not modified by the presence of strong gravitational
fields. In the following section we consider the case studied by
Mannheim \& Kazanas (1991) in which the scalar field {\it is} modified by
changes in the gravitational field.

We begin by applying Eq. (\ref{Lpsi}) to a binary system in circular motion
of radius $a$ and Keplerian angular frequency $\omega$. Using Kepler's
law $\omega^2 \, a^3 = GM$ we obtain in terms of $M_s = M / M_{\odot},
\mu_s  = \mu / M_{\odot}$, and $a_{10} = a / 10^{10}$ cm
\begin{equation}
L_{\Psi} = 8.7\times10^3 U_{DE} G^4M^2\mu^2/(c^7a^2) \simeq 10^6
(\mu_s \, M_s / a_{10})^2~~{\rm erg/s}
\end{equation}
where we have inserted needed factors of $c$ in Eq. (\ref{Lpsi}) $[(U_{DE}c)
(G^4/c^8)]$. Comparing with Eq.(1), $L_G \simeq 1.7 \times 10^{36} M_s^3
\mu_s^2 /a_{10}^5$ erg/s, we see that
\begin{equation}
\frac{L_{\Psi}}{L_G} = 10^{-30} (\frac{a_{10}^3}{M_s})
\label{sctograv}
\end{equation}
Thus quinton emission will
dominate emission of gravitational radiation only for very large values
of the orbit radius at which point they are both negligible.

Looking at larger mass objects, i.e. galaxies, clusters, superclusters etc...
we can set $M \sim (4/3) \pi a^3 \rho_c \Omega_M$ or $M_s \sim 10^{-32}
a_{10}^3$. Eq. (\ref{sctograv}) requires $a_{10} > 10^{11} M_S^{1/3}$ for
quinton production to win.
For two galaxies of $M \sim 10^{12} / M_{\odot}$, this gives $10^7$ light
years, somewhat larger that the average galaxy separation. For a star in a
galaxy circling the center of mass, $L_{\Psi}$ is negligible compared to $L_G$.

Going the other way, we can look at small systems, asking how much energy is
radiated in quintons in an excited
state lifetime of an atom. Going back to Eq. (\ref{Lpsi}) we insert $\omega a = v =
\alpha c/n_B$, where $n_B$ is the principal quantum number.  Taking $\mu$ to
be the mass of the electron, we see that
\begin{equation}
L_{\Psi} =  10^{-111} \, n_B^{-4}~~{\rm eV/s}
\label{lpsiatom}
\end{equation}

We can also ask for the enchancement that would result in the case that there
are large compact dimensions (Arkani-Hamed, Dimopoulos \& Dvali et al., 1998) and the quintons can travel in
the bulk. Following that reference
for the case $n=2$ of just two extra compact dimensions, we would get
from Eq. (\ref{psi}), for an atom ($a \sim 10^{-8}$),
\begin{equation}
\frac{d^3 x'}{x'^2} \rightarrow \frac{d^5 x'}{x'^4} \sim
\left(\frac{{\rm mm}}{a}\right)^2
\end{equation}
The result would be to modify Eq.(\ref{lpsiatom}) by a factor of $\sim 10^{28}$
which does not appear enough to make it of experimental interest.
There is  a larger effect if we consider a smaller (nuclear)
system such as $\alpha-$decay of a long-lived isotope. In that case,
a 10 MeV $\alpha-$particle could give $(\omega \, a)^2 \sim 10^{-2}
c^2$ so that Eq. (16) would be
\begin{equation}
L_{\Psi} =  10^{4}(10^3eV/cm^3) c \left(\frac{m_{\alpha}}{m_{Pl}}\right)^2
\left(\frac{\hbar c}{m_{Pl}}\right)^2 \left(\frac{\omega a}{c}\right)^4
\; (1 {\rm mm} /5 \cdot 10^{-13} {\rm cm})^4 \sim  10^{-45}~~{\rm eV/s}
\end{equation}
Thus, even a mole $(10^{24})$ of an isotope with a billion year half
life would have less than an electron volt of energy loss into
quinton radiation.
Finally, we turn to accelerator production, $p-p$ collisions at the
LHC.  We take $a \sim \sigma^{1/2} \sim 10^{-13}$ cm and compute the energy radiated into quintons in a collision.  \begin{equation}
\Delta E = L_{\Psi}\delta t
\end{equation}

with $\delta t = a/c$ and

\begin{equation}
L_{\Psi}=8.7\times10^3 (U_{DE}G^2)(\mu^2/\hbar) (a\omega/c)^4
(1mm/a)^4
\end{equation}

where we have, again, assumed two extra compact dimensions.  We have 
also assumed $a\omega=c$, although it is possible that, in a quantum 
treatment, we might have $a\omega\rightarrow aE$ giving a much larger 
result.  We have, of course, set $\mu = E$.  Again the result is 
small: $\Delta E \sim10^{-62}GeV$. Thus, based on a quinton field varying 
only over cosmological times, quinton radiation does not approach 
gravitational radiation for any of the 3 cases considered in atomic 
and nuclear transitions and high energy collisions, nor in the 
astrophysical binary systems considered.

\section{Quintessence Field Varying in Strong Gravitational
Field}

We turn now to the possibility that the scalar field is modified in the
presence of gravitational fields.  Specifically, we address the cases
of Section III in light of the results of the work of Mannheim \& Kazanas 
(1991). These authors considered Eq. (3) written in the form
\begin{equation}
\frac{1}{(-g)^{1/2}} \partial_{\mu}[(-g)^{1/2}\partial^{\mu} \Psi] =
\frac{\partial V}{\partial \Psi} \simeq 0
\label{scalf}
\end{equation}
along with Einstein's equations
written in the form
\begin{equation}
\frac{1}{8 \pi G} G^{\mu \nu}-  \partial^{\mu}\Psi \partial^{\nu} \Psi
+ \frac{1}{2} g^{\mu \nu}\partial^{\alpha}\Psi \partial_{\alpha} \Psi
 = - g^{\mu \nu}V(\Psi)
\label{einstein}
\end{equation}
They find, as did Buchdahl (1959), closed form solutions for the case
$V(\Psi)=0$. With $V(\Psi) \simeq m^2_{\rm eff} \Psi^2$ and $m_{\rm eff}
\sim 10^{-33}$ eV, as demanded by the dark energy observations, their
two solutions should be good approximations.  One is that $\Psi$ is a constant, unaffected by gravitational
fields.  This, of course, is just the case considered above.  The second
solution, of the coupled equations, is
\begin{equation}
ds^2 = -H(\rho) dt^2 + J(\rho)[d \rho^2 + \rho^2 d\Omega]
\end{equation}
The functions in the above equation
are
\begin{eqnarray}
\Psi (\rho) & = & \frac{K}{2 r_0}ln \left(\frac{\rho - r_0}{\rho + r_0}\right) + constant
\\
H(\rho) & = & \left(\frac{\rho - r_0}{\rho + r_0}\right)^{-d/2 r_0}\\
J(\rho) & = & \left(1 - \frac{r_0^2}{\rho^2}\right)^2 \left(\frac{\rho - r_0}
{\rho + r_0}\right)^{-d/2 r_0}
\end{eqnarray}
where
\begin{equation}
d = 2MG = 4(r_0^2 -  \pi G K^2)^{1/2}
\end{equation}
and it can be shown (Mannheim \& Kazanas 1991) that $r_0$ is 
restricted to the region
\begin{equation}
MG/2 \le r_0 \le MG,
\end{equation}
indicating that $r_0$ is of the same order of magnitude as the 
Schwarzschild radius. We set $\pi K^2 = GM^2 \beta^2, ~r_0 =
MG\gamma$ with $0 \le \beta \le 1$ and $1/2 \le \gamma \le 1$.  We assume
that the dimensionless quantities, $\beta$ and $\gamma$, are independent
of the gravitational field, that is, like Newton's constant, are the
same for all masses.

Using the above equations we compute $\Gamma^{\mu}, \Psi_R, L_{\Psi}$
and $L_{\Psi}/L_{GW}$. We first obtain
\begin{eqnarray}
g^{-1/2} \partial_{\rho}(g^{1/2} g^{\rho \rho}) &=& \Gamma^{\rho}\\
&=& \frac{2}{\rho^3}\left(1 - \frac{r_0^2}{\rho^2}\right)^2 \left(\frac{\rho -
r_0}
{\rho + r_0}\right)^{-1/2 \gamma}(\rho^2 - 6 r_0^2 - \frac{1}{\gamma}r_0 \rho)\\
& \simeq & \frac{2}{\rho} ~~{\rm for} ~~\rho > a \gg  r_0
\end{eqnarray}

Similarly, we have $\Gamma^0 = (2/\rho)\partial_0\rho, ~\partial_{\rho}
\Psi \simeq K/\rho^2$, and $\partial_0 \Psi \simeq (K/\rho^2)\partial_0\rho$.

We rewrite the solution of the wave equation (3) as
\begin{equation}
\Psi_{MK}(\vect{x},t) = x^{-1} \int d^3x' \Gamma^{\mu}(\vect{x'},t')
\partial_{\mu}\Psi_0
\end{equation}
for each of the two bodies in the binary system separately.  Using $\vect{\rho} =
\vect{x'-a_i(t_i)}$, we will add the two contributions to $\Psi_{MK}$.  The time and
$\rho$ components give much different results for  $\Psi_{MK}$ and $L_{MK}$.
For the time component, $\partial_0\vect{\rho}$ is simply $\partial_0 \vect{a}$
and the leading contribution is
\begin{equation}
\Psi_{MK,0} = x^{-1}\int d^3x' (4i\omega\vect{a_1}\cdot\hat{x'})
(2i \omega \vect{a_1} \cdot\hat{x'}) K_1/x'^3 = (32 \pi /3x) e^{2i\omega(t-x)}
K_1 a_1 ^2 \omega^2 ln(1/ a\omega) + (1 \rightarrow 2)
\end{equation}
which gives, after using $\pi(K_1a_1^2+K_2a_2^2)^2 = \beta^2G\mu^2a^4$,
\begin{equation}
L_{MK,0} = (64\pi/3)^2 \beta^2 (G\mu^2) (a\omega)^4 \omega^2 ln^2(a\omega/c)
\label{LMK0}
\end{equation}
The $\rho$ contribution is given by
\begin{equation}
\Psi_{MK,\rho} = (2 K_1/x)\int d^3 x' /\vert\vect{x'}-\vect{a_1}\vert^3 + (1 \rightarrow 2)
\end{equation}
We expand the denominator(s) in powers of the $a_i$; noting that the 
term of first order in $a_i$ vanishes after the angular integration, we
are left with terms of order 2.  The result is
\begin{equation}
\Psi_{MK,\rho} = (6K_1/x)\int d^3x' [4(\hat{x'}\cdot\vect{a_1})^2-a_2^2]/x'^5
+(1\rightarrow2) = 2 \pi (K_1+K_2)
\end{equation}
from which we get
\begin{equation}
L_{MK,\rho} = 2(4\pi)^2 G \mu^2 \omega^2 \beta^2
\label{LMK}
\end{equation}
Here, we have approximated $\pi (K_1^2 + K_2^2)$ by its equal mass value.

Clearly, the relative factor of $(a\omega/c)^4$ between $L_{MK,0}$ and
$L_{MK,\rho}$ means that the latter will be more important for astrophysical
objects and as well as atomic and nuclear ones, while the former will dominate in
high energy physics.  Turning first to astrophysics, one may note

\begin{equation}
\frac{L_{MK,\rho}}{L_{GW}} \simeq \frac{16 \pi \beta^2 G \mu^2
\omega^2}{\frac{64}{5}
G \mu^2 a^4 \omega^6} = \frac{5 \beta^2 /4}{a^4 \omega^4} \simeq \frac{\beta^2}
{(v/c)^4}
\end{equation}
where $v$ is the velocity of the lighter member in the non-equal mass
case. We note the similarity between Eq.(\ref{LMK}) and $L_{GW}$. This might
be
expected since there is no additional mass scale, such as the
$U_{DE}$ in section II which enters in forming $\Psi_R$. .

We can now estimate an upper bound on $\beta$ from the Hulse -- Taylor
pulsar which has an orbital period of 28,000 sec and the fact that 
the rate change of its orbital period agrees to better than one percent
with the prediction of General Relativity. From the fact that
$L_{MK}/L_{GW} < 0.01$ we obtain that $\beta^2 < 2 \times 10^{-3} (v/c)^4
\simeq 10^{-15}$. While this appears a stringent bound on the constant
$K$, understanding its full implications awaits simultaneous solution of
Equations (\ref{scalf}, \ref{einstein}) in the interior region as well as the exterior
region which should permit evaluating $K$ and $r_0$ in terms of the interior
mass distribution.

Turning to the high energy case, we compute the energy that would be 
lost into quinton radiation in proton-proton collisions.  $L_{MK,0}$ of 
Equation (\ref{LMK0}) above dominates.  In it, we take $\mu$ to be 7 
TeV as with the LHC in section 3 and $\omega$ to be $c/a$, while 
recognizing that it might be larger in a quantum treatment.  Again, we 
multiply $L$ by $a/c$ to obtain the energy radiated during the encounter.  
We include enhancement from two extra compact dimensions.  We ignore 
the logarithm. These give
\begin{equation}
\Delta E = (64\pi/3)^2 (E/m_{Pl})^2 \beta^2 
(\omega^2a/c) (1mm/a)^4 = 10^{18}\beta^2 GeV
\end{equation}
This implies that, if significant unexplained energy loss in $p-p$ collisions at the LHC is not found, a bound on $\beta^2$ slightly better than that above from the Hulse-Taylor pulsar can be inferred for models with compact extra dimensions.  It should also be possible to infer such a bound from cosmic ray data from the Auger project (www.auger.org) which will study p-p scattering at about 30 times LHC energies (at the center of Mmss).  One might even infer from the existence of ultra high energy cosmic ray observations that proton-proton interactions do not lose large amounts of energy into unobserved particles, implying a bound on $\beta^2$ of about $10^{-21}$. On the other 
hand, the cosmic ray spectrum does exhibit a feature known as the ``knee" 
at an energy $E_k \simeq 10^{15.5}$ eV which  corresponds to roughly 1 TeV 
at the center of mass. This is a steepening in the slope cosmic ray spectrum by 
0.3 - 0.4 over half a decade in energy. The very limited energy range over which 
this change in the cosmic ray spectral index occurs essentially precludes its 
explanation as simply a cosmic ray propagation effect. In fact Kazanas \& 
Nicolaidis (2003) suggested that this feature heralds the emergence of physics 
beyond the Standard Model. In particular, these authors have argued that such a 
feature is the result of energy lost in cosmic ray collisions to gravitons, 
as suggested by the theories of extra large dimensions which presumably 
have the graviational constant increase with energy to that of strong 
interactions at energies $\sim 1$ TeV. It would not be in conflict with 
our Hulse-Taylor pulsar bound, if quinton emission were also to contribute 
to this cosmic ray feature.

\section{Summary}

General relativity requires that any quintessence field couple, through the
covariant derivative, to the gravitational field.  Thus, any time varying
gravitational field will produce quintons.  The rate of production, however, will be
proportional to the square of the derivative of the quintessence field.  We have
evaluated that rate, in a classical approximation, in Section II, for the case in
which the field only varies over cosmological times.  We
applied the results to a variety of binary systems in Section III, but found no
cases of interest in which energy loss through quintons dominated energy loss
through gravitons.

In Section IV we turned to the perhaps more realistic case in which the space
variation of the quintessence field is affected by the presence of a gravitational
field.  We worked with the exact (exterior) solution for the massless case as
written down by Mannheim and Kazanas (1991).  The result, in that case, was
quinton production dominated by different components for low velocities of the binary system than for high.  We were
able to bound an integration constant in the solution cited by requiring that
quinton emission from the Hulse-Taylor binary pulsar system represents less
than one percent of the energy loss.  That limit on the parameter was sufficient to
make it difficult to identify observable effects in the astrophysical phenomena considered.  We were also able to show that, with the assumption of large, compact extra dimensions similar and stronger bounds could be derived from data, when available, from the LHC and project Auger (and perhaps from current cosmic ray data showing the existence of ultra high energy cosmic rays).

In summary, it appears that quinton emission is unlikely to be of importance in
interpreting signals received by LIGO or LISA, or in laboratory experiments if the
quinton field has no coupling to matter beyond the indirect coupling provided by
the covariant derivative or if large compact extra dimensions do not exist.

\begin{acknowledgments}
It is a pleasure to acknowledge helpful conversations with Richard Fahey, 
Breno Imbiriba, Doris Rosenbaum and Michael Turner.
\end{acknowledgments}

\clearpage
\end{document}